\documentclass[aps,prd,twocolumn,showpacs,amsmath,amssymb,nofootinbib]{revtex4-1}

\usepackage{graphicx}
\usepackage[english]{babel}
\usepackage{color}

\DeclareMathOperator{\sh}{\mathrm{sh}}
\DeclareMathOperator{\ch}{\mathrm{ch}}
\DeclareMathOperator{\tnh}{\mathrm{th}}

\newcommand{\ph}{\varphi}

\newcommand{\tens}[1]{{#1}}
\newcommand{\grad}{{\tens{d}}}

\newcommand{\Ind}{{\scriptstyle\mathrm{Ind}}}
\newcommand{\Lob}{{\scriptscriptstyle\mathrm{Lob}}}

\newcommand{\crit}{{\scriptstyle\mathrm{cr}}}
\newcommand{\mx}{{\scriptstyle\mathrm{max}}}
\newcommand{\plane}{{\scriptstyle\mathrm{hp}}}

\newcommand{\ren}{{\scriptstyle\mathrm{ren}}}

\newcommand{\scri}{{\mathcal{I}}}

\newcommand{\realn}{{\mathbb{R}}}

\newcommand{\xP}{{\bar{x}}}
\newcommand{\yP}{{\bar{y}}}
\newcommand{\zP}{{\bar{z}}}
\newcommand{\rP}{{\bar{r}}}

\renewcommand{\tens}[1]{{\boldsymbol{#1}}}

\newcommand{\jpgimgdir}{}
\newcommand{\pdfimgdir}{}

\begin{document}
\title{Entanglement entropy of spherical domains in anti-de~Sitter space}

\author{Pavel Krtou\v{s}}

\email{Pavel.Krtous@utf.mff.cuni.cz}

\affiliation{
ITP,
Faculty of Mathematics and Physics, Charles University in Prague,
V~Hole\v{s}ovi\v{c}k\'ach 2, Prague, Czech Republic}

\author{Andrei Zelnikov}
\email{zelnikov@ualberta.ca}

\affiliation{TPI, Department of Physics, University of
Alberta, Edmonton, Alberta T6G 2E1, Canada}

\date{November 7, 2013}  

\begin{abstract}
It was proposed by Ryu and Takayanagi that the entanglement entropy in
conformal field theory (CFT) is related through the AdS/CFT correspondence to the
area of a minimal surface in the bulk. We apply this holographic
geometrical method of calculating the entanglement entropy to study the
vacuum case of a CFT which is holographically dual to empty anti-de~Sitter (AdS)
spacetime. We present all possible minimal surfaces spanned on one or two
spherical boundaries at AdS infinity. We give exact analytical
expressions for the regularized areas of these surfaces
and identify finite renormalized quantities. In the case of two disjoint
boundaries the existence of two different phases of the entanglement
entropy is confirmed \cite{Klebanov:2007ws,Lewkowycz:2012mw}. A trivial phase
corresponds to two disconnected minimal surfaces, while the other one corresponds to
a tube connecting the spherical boundaries. A transition between
these phases is reminiscent of the finite temperature deconfinement
transition in the CFT on the boundary. The exact
analytical results are thus consistent with previous numerical and approximate
computations. We also briefly discuss the character of a spacetime extension of
the minimal surface spanned on two uniformly accelerated boundaries.
\end{abstract}


\maketitle

\section*{Introduction}
\label{sc:intro}

The famous Bekenstein--Hawking area law
\cite{Bekenstein:1972tm,Hawking:1971vc}
\begin{equation}\label{Bekenstein-Hawking}
S_{BH}= {k_{\scriptscriptstyle B} c^3\over \hbar}{{A}\over 4\, G}
\end{equation}
for the entropy of black holes connects thermodynamics,
gravity and relativistic quantum field theory. This relation remains valid not
only in Einstein's gravity in four dimensions but in higher dimensions too,
as long as the gravitational constant $G$ is ${D}$-dimensional and the area
${A}$ is understood as the volume of the \mbox{$(D{-}2)$-dimensional} horizon surface.

In quantum field theory (QFT) an entanglement entropy can be attributed
to any surface formally dividing the system in two parts. The leading UV
contribution of quantum fields to the entanglement entropy is proportional to the
area of the dividing surface \cite{Bombelli:1986rw,Frolov:1993ym,Srednicki:1993im}.
This property is strikingly similar the Bekenstein--Hawking area law.
The analogy with black hole entropy is not accidental.
One may consider a black hole horizon as a surface separating the interior
of the black hole from its exterior. To define a wave function of all quantum
fields
in the black hole spacetime \cite{Barvinsky:1994jca} one can use
an analogue to the Hartle-Hawking no-boundary proposal. Using this wavefunction one can construct the density matrix for
the fields inside the black hole and derive the corresponding entanglement
entropy, which is proportional to the area of the horizon, but the coefficient
of
proportionality formally diverges. However, one has to take into account that
quantum fields on a curved background spacetime also contribute to the effective
gravitational constant. It is amazing that
quantum contributions to the entropy per unit area of a horizon are described by
the same functions as quantum corrections to the gravitational coupling
\cite{Susskind:1994sm}. The interpretation of the Bekenstein--Hawking formula
as the entanglement entropy becomes even more striking
in the framework of induced gravity models
\cite{Sakharov:1967pk}
where the gravitational coupling is completely defined by quantum field
contributions. In these models the leading UV contribution to the entanglement
entropy of the horizon
\cite{Jacobson:1994iw,Frolov:1996aj,Frolov:1996qh,Frolov:2003ed} is given by the
formula $A/(4G_\Ind)$  and is finite as soon as the induced
gravitational constant $G_{\Ind}$ is finite.
It was also proposed \cite{Barvinsky:1994jca}  that in generic static
spacetimes with horizons,
the minimal area surface inside the slice of a constant time may
play an important role in the definition of the entanglement entropy of a black
hole.

Recently, holographic computations of the entanglement entropy in conformal
field theory (CFT) at infinity of the anti-de~Sitter (AdS)
spacetime have seen a lot of attention and development.
The original conjecture for entanglement entropy by Ryu and Takayanagi
\cite{Ryu:2006bv,Ryu:2006ef,Nishioka:2009un}
is that in a static configuration the entanglement entropy of a subsystem
localized in a domain $\Omega$ is given by the formula\footnote{From now on we
use a
$k_{\scriptscriptstyle B}=c=\hbar=G=1$ system of units.}
     \begin{equation}\label{Ryu-Takayanagi}
     S_{\Omega}={{A}_{\Sigma_\Omega}\over 4 \,G} .
     \end{equation}
Given a static time slice
(the \mbox{${(D{-}1)}$-dimensional} bulk space), the
\mbox{$({D{-}2})$-dimensional}
domain ${\Omega}$ belongs to an infinite boundary ${\mathcal{I}}$ of the bulk and
the area ${A_{\Sigma_\Omega}}$ in Eq.~\eqref{Ryu-Takayanagi} is to be understood as
the area of a ${(D{-}2)}$-dimensional minimal surface $\Sigma_{\Omega}$
in the bulk spanned on the boundary ${\partial \Omega}$ of the subsystem
(i.e., ${\partial\Sigma_{\Omega}=\partial \Omega}$).

The holographic derivation of the Ryu--Takayanagi formula for the entanglement entropy
was proposed in \cite{Fursaev:2006ih} using the replica trick. In the replica method there naturally appears a more
general notion of the Renyi entanglement entropy. But the QFT derivation of the
Ryu--Takayanagi relation based on the calculation of Renyi entropies requires a
different approach \cite{Hartman:2013mia,Faulkner:2013yia}.
In QFT with gravity duals, formula \eqref{Ryu-Takayanagi} was proven for
$\text{AdS}_3$ \cite{Hartman:2013mia,Faulkner:2013yia}. In a more general
case of Euclidean gravity solutions without Killing vectors, arguments
supporting the validity of the Ryu-Takayanagi formula were given in
\cite{Lewkowycz:2013nqa,Faulkner:2013ana}.
In the last few years the conjecture by Ryu and Takayanagi has
been generalized to gravity theories with higher curvature interactions
\cite{Hung:2011xb,Casini:2011kv,Myers:2013lva,Bhattacharyya:2013jma} or some
other deformations of the gravity theory \cite{Hung:2011ta}.
An excellent up-to-date review of the
entanglement entropy and black holes can be found in \cite{Solodukhin:2011gn}.

Calculation of entanglement entropy for a subsystem localized in two
disjoint regions is particularly interesting, since it can be used as a probe
of confinement \cite{Klebanov:2007ws,Lewkowycz:2012mw}. It was demonstrated
\cite{Klebanov:2007ws,Hirata:2006jx} that in confining backgrounds there
are generally more solutions for minimal surfaces in the bulk spanned
on the boundaries of these disjoint regions. However, there is a maximum
distance between the regions beyond which the tube-like minimal surface
connecting both components ceases to exist. There is also a critical
scale beyond which a solution with disconnected minimal surfaces dominates over
the
connected one. In the QFT language this critical behavior is analogous to
a deconfinement transition at a finite temperature.

In the case of a few disjoint
regions the minimal surfaces connecting their boundaries are generally not
unique and their areas differ. The conventional wisdom is that the entanglement
entropy is related to the surfaces of minimum area. This choice
guarantees the required strong subadditivity property \cite{Headrick:2007km} of
the entanglement entropy. There were some proposals \cite{Hubeny:2007re} how to modify
this ``least area'' rule while still satisfying the strong
subadditivity property.

In this paper we study minimal surfaces in the pure AdS spacetime. We
show that many properties of the entanglement entropy, such as the critical
behavior \cite{Klebanov:2007ws} demonstrated for the asymptotically AdS
spacetimes with black holes in the bulk, exist already in the pure AdS.

The main result is that we are able to find exact solutions for all minimal
surfaces spanned on one or two spherical boundaries positioned arbitrarily
at conformal infinity ${\scri}$. In this short paper we give
analytical formulas for the regularized and renormalized area of these
minimal surfaces. The explicit form of the surfaces and its derivation is
presented in a more detailed paper \cite{KrtousZelnikov:minsurf}. We also shortly
discuss the spacetime character of a minimal surface spanned on two accelerated
spherical domains.

\section*{Spherical boundaries at infinity}
\label{sc:sphbnd}

We start with geometrical preliminaries concerning the bulk space and
with the characterization of the spherical domains at infinity.
We will discuss only a \mbox{${3{+}1}$-dimensional} AdS spacetime although most of the discussion can be extended to higher dimensions.

The AdS spacetime has many Killing symmetries and can be viewed as a
static spacetime in various ways. However, in all cases the spatial section---the bulk space---has
the hyperbolic geometry of Lobachevsky space.
To describe it, we use cylindrical coordinates
${\rho,\,\zeta,\,\ph}$ and Poincar{\'e} coordinates ${\xP,\,\yP,\,\zP}$ in which
the metric reads
\begin{equation}\label{Lobmtrccyl}
    \frac1{\ell^2}\,\tens{g}_{\Lob}
    {=} \grad \rho^2 {+} \ch^2\!\rho\,\grad\zeta^2{+}\sh^2\!\rho\,\grad\ph^2
    {=} \frac{1}{\zP^2}\,\bigl(\grad \xP^2 {+}\grad \yP^2{+}\grad\zP^2\bigr)\,.
\end{equation}
Here, $\ell$ is the characteristic scale describing the radius of curvature of
AdS, as well as of its spatial section.
The coordinates are related by
${\zP=\rP/\ch\rho}$,
${\xP=\rP\tnh\rho\cos\ph}$,
${\yP=\rP\tnh\rho\sin\ph}$, with
${\rP = \exp\zeta}$.

The conformal infinity ${\scri}$ of the spatial section is the conformal
sphere. In
cylindrical coordinates it is given by ${\rho\to\infty}$,
${\zeta\to\pm\infty}$. In Poincar{\'e} coordinates it is represented by the
plane ${\zP=0}$ plus one improper point ${\rP\to\infty}$.

\begin{figure}[b]
\includegraphics[width=1.67in]{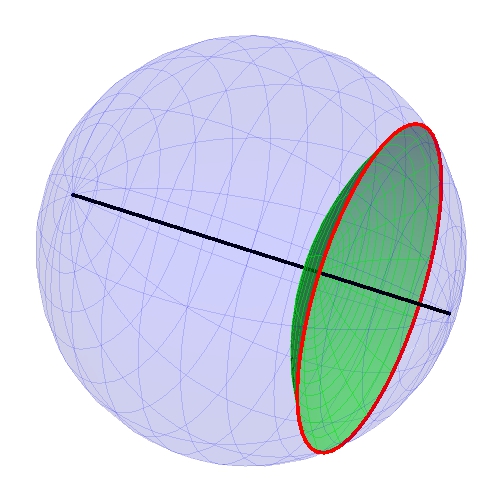}\hfill
\includegraphics[width=1.67in]{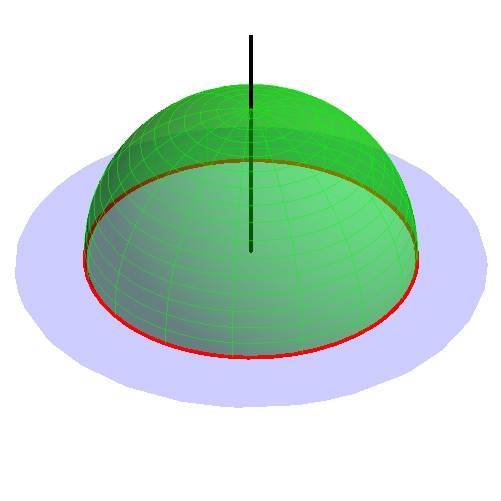}\\[1ex]
\caption{\label{fig:hyperplane}%
\textbf{Hyperplane corresponding to the circular boundary at infinity.}
The hyperplane (the hypersurface of zero extrinsic curvature) reaches infinity
at the circular boundary which divides infinity into two domains.
The Poincar\'e spherical (left) and spherical half-space
(right) representation of hyperbolic space is shown.}
\end{figure}

By the \emph{circular} (in higher dimension, \emph{spherical}) boundary
${\partial\Omega}$ of a ball-like domain ${\Omega}$ at infinity ${\scri}$ we
mean a \mbox{${1}$-dimensional} surface given by infinite points of a
\mbox{${2}$-dimensional} hyperplane in the bulk (hyperplane in the sense of
hyperbolic geometry, a hypersurface with zero extrinsic curvature). The
circular boundaries at infinity are thus in one-to-one
correspondence with hyperplanes in the bulk. Visualization of such a hyperplane
and the corresponding circular boundary can be found in Fig.~\ref{fig:hyperplane}.
The boundary ${\partial\Omega}$ can be understood also as the boundary of the
complementary domain ${\scri\backslash\Omega}$.

From a point of view of the conformally spherical geometry on ${\scri}$, all
such boundaries are equivalent. It is a reflection of the trivial fact that all
hyperplanes in the bulk are isometric. Therefore we do not have any quantity
measuring a `size' of spherical boundaries at infinity.

However, in many calculations, both in the bulk or at infinity, we need to
regularize various quantities. Instead of working at ${\scri}$ we restrict on
some cut-off surface at large finite size. Then we can measure the sizea of
the circular boundaries using the geometry on the cut-off surface. But, since
the
choice of the cut-off can be rather arbitrary, the regularized size of the
spherical boundary can be only an intermediate quantity,
and physically measurable quantities should be cut-off independent.

Two circular boundaries can be in three qualitatively distinct positions: (i)
disjoint boundaries (corresponding hyperplanes are ultraparallel),
\footnote{
Let us note that two disjoint circles positioned `side by side' or `one inside
of
another' in the Poincar{\'e} planar representation of infinity are equivalent;
they
differ only by a choice of the improper point which closes planar part of
infinity into sphere.}
(ii) boundaries crossing each other (the hyperplanes intersect in a line), and
(iii) boundaries touching in one point (the corresponding hyperplanes are
asymptotic).

In the first case we can define the distance of the boundaries as a distance of
the corresponding hyperplanes.
To the crossing circular boundaries we can assign an angle of the corresponding
hyperplanes. Finally, all pairs of touching boundaries are equivalent. Indeed,
all pairs of asymptotic hyperplanes in an arbitrary position are isometric to
each other. In global hyperbolic space there is no measure which could
distinguish them.

\section*{Surface spanned on one boundary}

Now we review known results for a minimal surface spanned on one circular
boundary.
Such a minimal surface is trivial: it is the hyperplane which defines the
boundary. If we choose the axis of the cylindrical coordinates perpendicular to
the hyperplane, the hyperplane is given by ${\zeta=\text{const}}$. If we choose the axis
inside the hyperplane, the hyperplane is given by ${\ph=\ph_0,\,\ph_0+\pi}$. In
the Poincar{\'e} coordinates the hyperplane is represented as a plane
orthogonal to
the infinity surface ${z=0}$ or as a hemisphere with the center at ${z=0}$ (here
we used a language of the conformally related Euclidian geometry with Cartesian
coordinates ${\xP,\,\yP,\,\zP}$).

To demonstrate different regularizations used later, we can write down the area
of a hyperplane measured up to a cut-off. For the hyperplane othogonal to the
axis we have
\begin{equation}\label{rotsym-planearea}
  A_{\plane}=2\pi\ell^2(\sqrt{1+P^2}-1)
    =C\ell\Bigl[1-\frac1P+\mathcal{O}\Bigl({\frac1{P^2}}\Bigr)\Bigr]\;,
\end{equation}
where ${C=2\pi\ell P}$, with ${P=\sh\rho_*}$, is the circumference of the circular boundary on the cut-off surface ${\rho=\rho_*\gg1}$.

For the hyperplane which contains the axis we have
\begin{equation}\label{trsym-planearea}
  A_{\plane}=2L\;\ell\sqrt{1-Z^{-2}} = 2L\ell \Bigl[1+\mathcal{O}\Bigl({\frac1{Z^2}}\Bigr)\Bigr]\;,
\end{equation}
where ${L=Z\Delta\zeta_*\ell}$, with ${Z=\ch\rho_*}$, is the length of the
boundary at the cut-off surface. In this case, the circular boundary is
represented by two lines ${\rho=\infty}$, ${\zeta\in\realn}$, ${\ph=\ph_0,\,\ph_0+\pi}$.
We thus have to introduce two cut-offs: an ultraviolet one,
${\rho=\rho_*}$, in the direction away from the axis, and
an infrared\footnote{The distinction between ultraviolet (UV) and infrared (IR)
cut-off is more or less conventional here. The cut-off labeling the regularized
surface near infinity is called UV since it corrresponds to a UV cut-off in the
related CFT. The IR cut-off is an extensive one; it corresponds to the length
along a translation symmetry.}  one, ${\zeta=\pm\Delta\zeta_*/2}$, along the axis.

Finally, the area of the hyperplane represented by a half-plane in
the Poincar{\'e}
coordinates, say ${\xP=\text{const}}$, is
\begin{equation}\label{horsym-plane}
    A_\plane = L\ell\;.
\end{equation}
${L={\Delta\yP_*}/{\zP_*}}$ is again the length of the circular
boundary at the cut-off surface ${\zP=\zP_*\ll1}$. It is also infrared
divergent: one has to cut-off the ${\yP}$ direction at ${\yP=\pm\Delta\yP_*/2}$.

In all three cases we recognize the well-known property that the leading
diverging
term of the minimal surface is (up to a constant scale) given by the regularized
size of the boundary at infinity. Clearly, the exact expression for the
divergent term depends on the regularization scheme, however, in all cases it
can be interpreted as the regularized size of the boundary at infinity
\cite{Hirata:2006jx,Myers:2010tj,Takayanagi:2012kg}.

The area of the trivial minimal surface spanned on one circular boundary
can be used to eliminate the infinite contributions to the area for more
complicated surfaces. We define the renormalized area of a surface by
subtracting the area of hyperplanes spanned on the same boundaries at infinity.
In this sense, the trivial minimal surface has vanishing renormalized area.

\section*{Surfaces spanned on two boundaries}

\begin{figure*}
\includegraphics[width=2.1in]{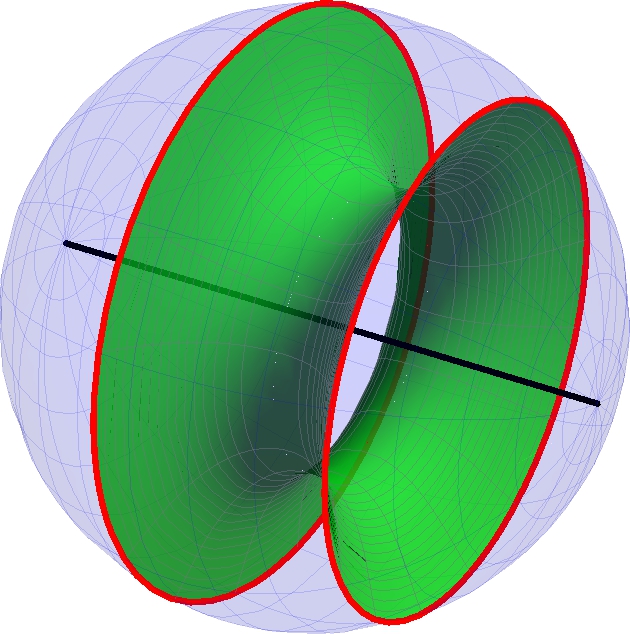}\hfill
\includegraphics[width=2.1in]{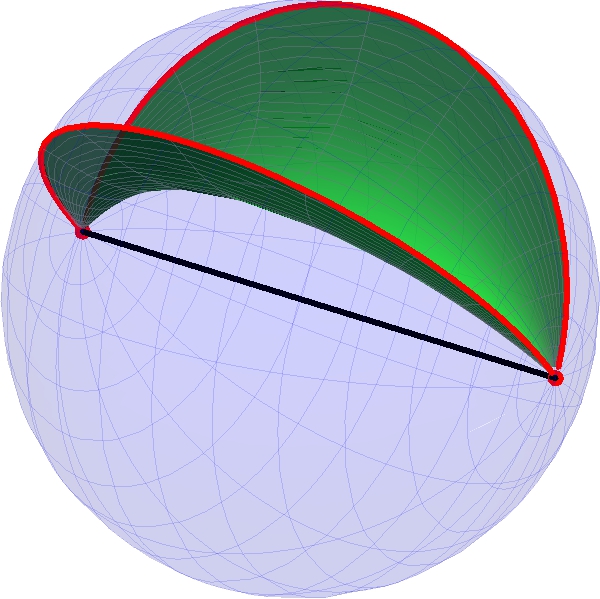}\hfill
\includegraphics[width=2.1in]{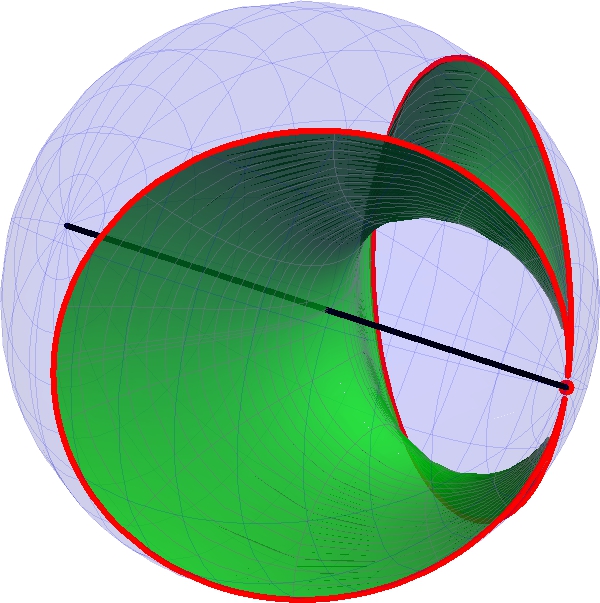}\\[2ex]
\includegraphics[width=2.1in]{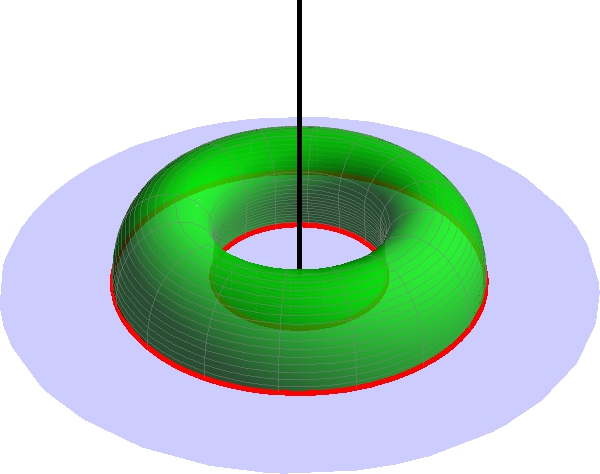}\hfill
\includegraphics[width=2.1in]{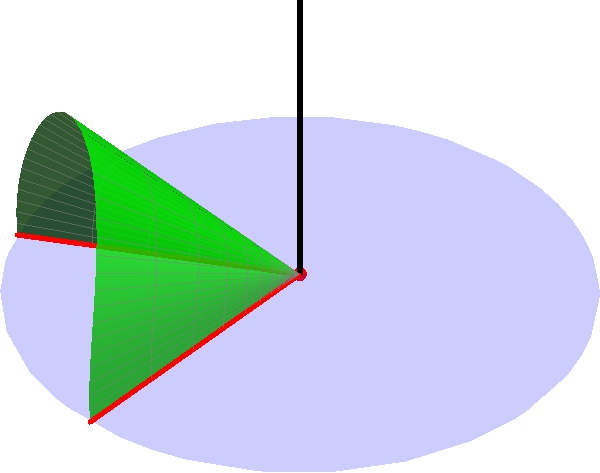}\hfill
\includegraphics[width=2.1in]{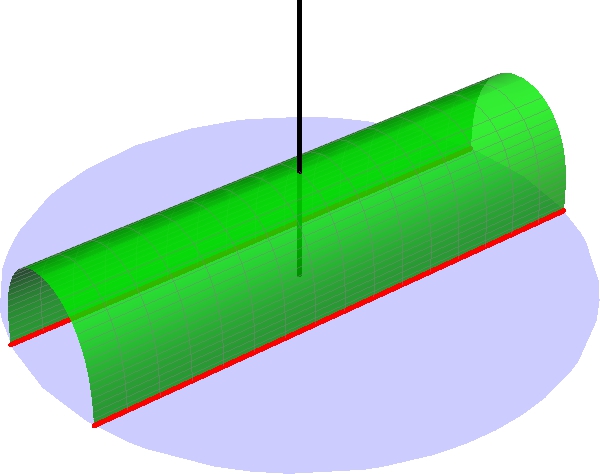}\\[1ex]
{\small (a)\hspace{18em}(b)\hspace{18em}(c)}\\
\caption{\label{fig:ms}%
\textbf{Minimal surfaces spanned on circular boundaries.} The surfaces are
visualized in Poincar{\'e} spherical (top) and Poincar{\'e} half-space (bottom)
models. (a) The tube-like surface spanned on two disjoint boundaries. (b) The
surface spanned on two semicircles joining opposite poles. (c) The surfaces
spanned on two touching circles.}
\end{figure*}

\begin{figure*}
\includegraphics[width=2.2in]{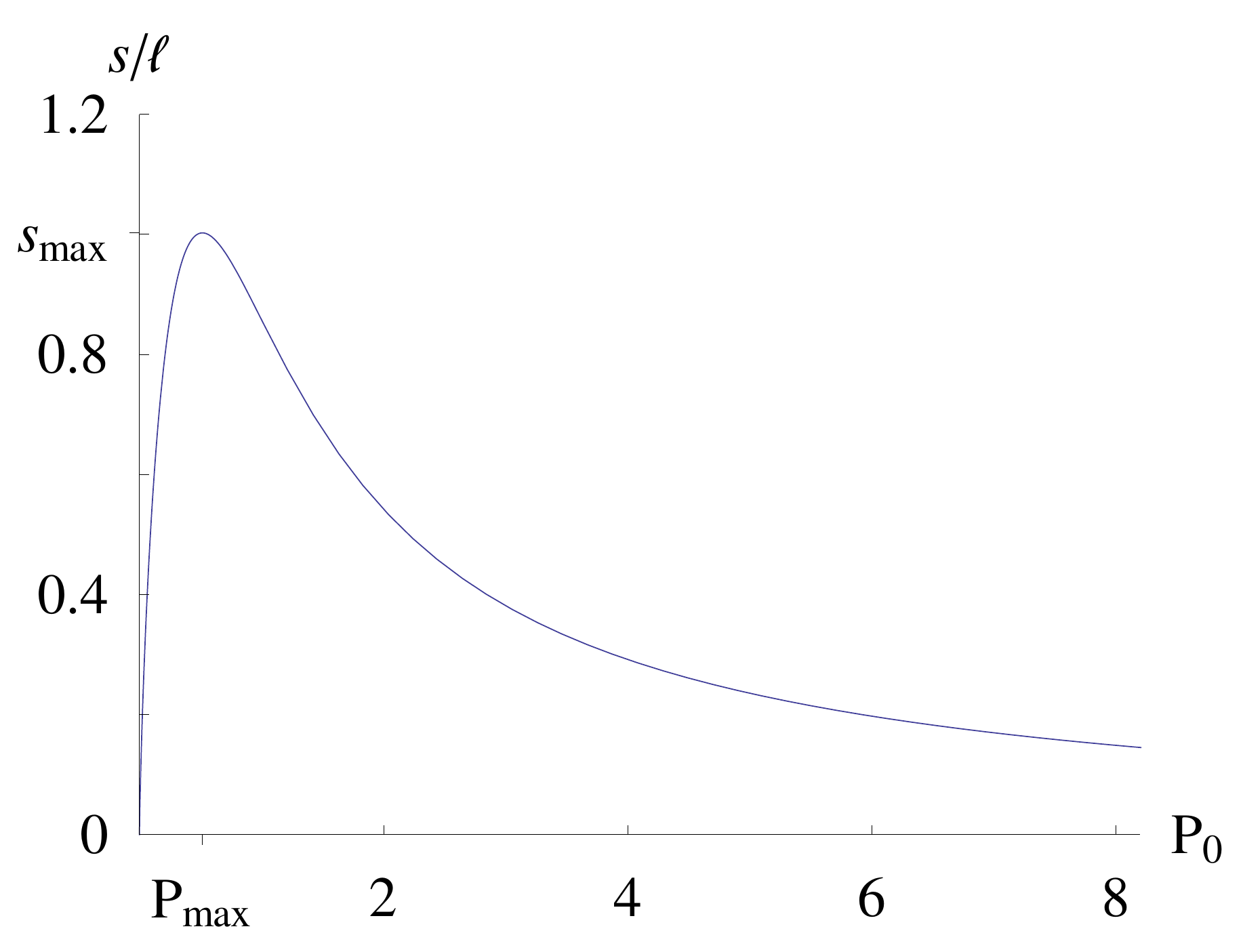}\hfill
\includegraphics[width=2.2in]{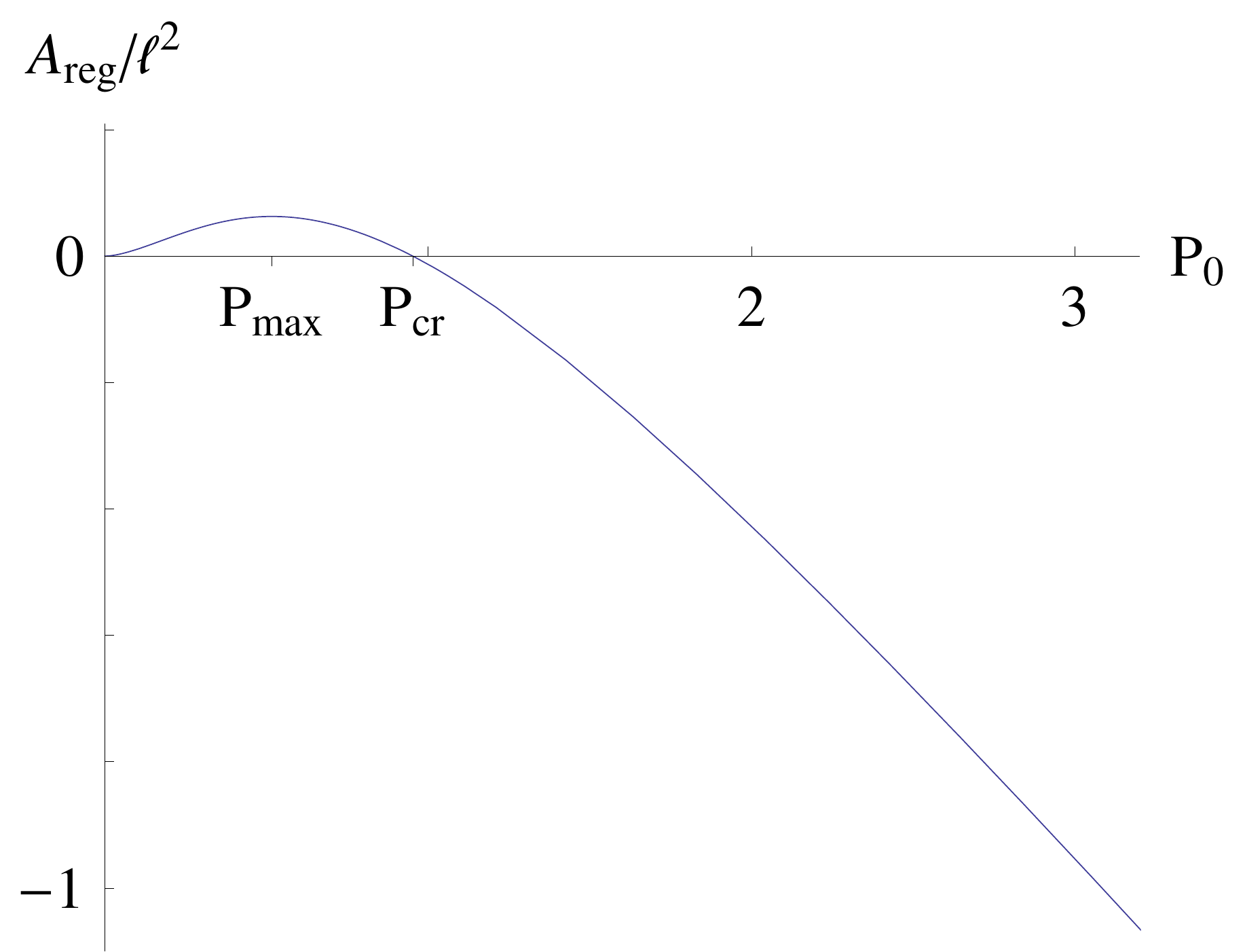}\hfill
\includegraphics[width=2.2in]{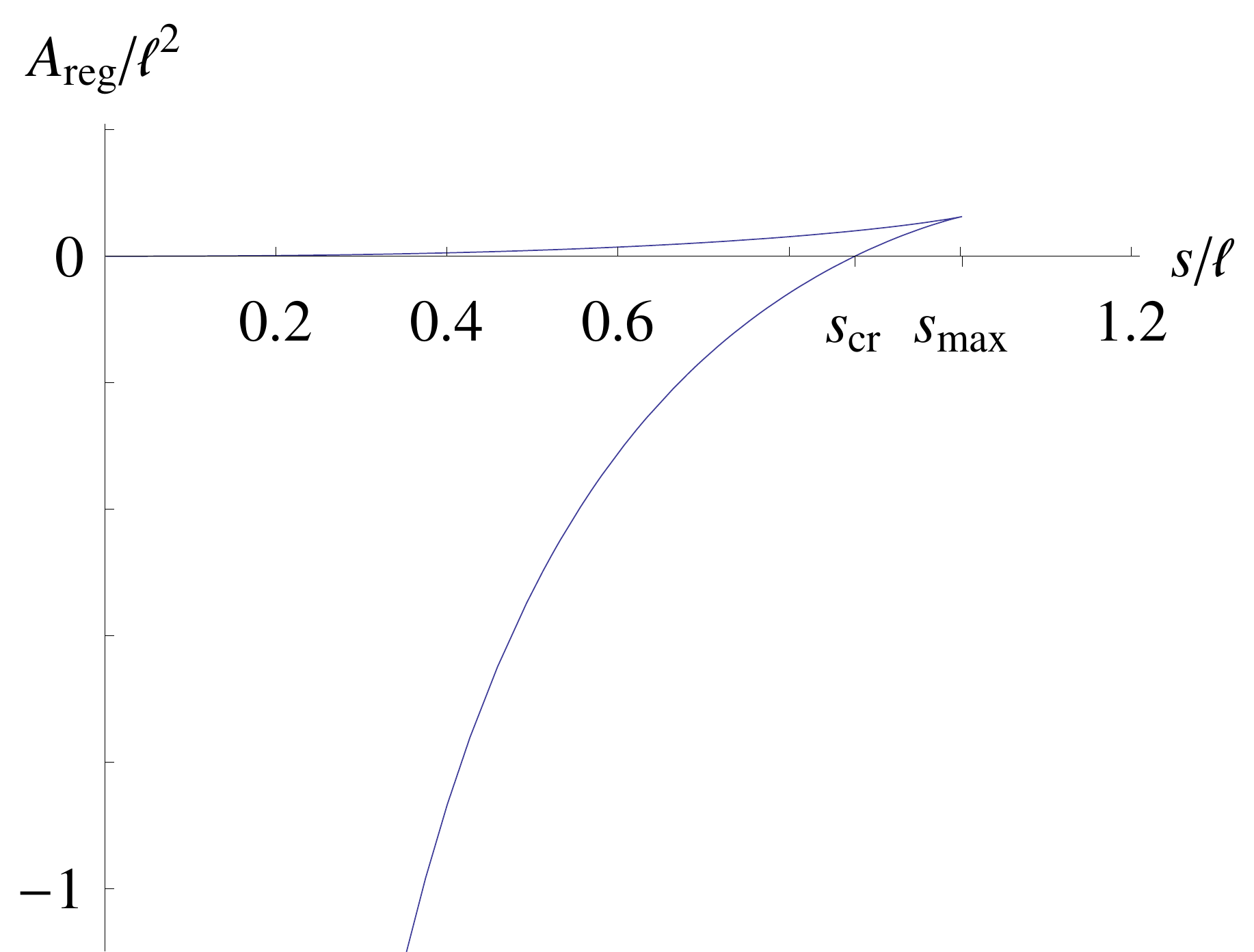}\\[1ex]
{\small (a)\hspace{18em}(b)\hspace{18em}(c)}\\
\caption{\label{fig:area}%
\textbf{Renormalized area of the minimal surface spanned on two boundaries.} (a) Relation between the distance ${s}$ of the boundaries and the closest approach ${P_0}$ of the tube to the axis. (b) Renormalized area of the tube as a function of ${P_0}$. (c) Renormalized area as a function of the distance ${s}$.}
\end{figure*}

\bigskip\noindent\textbf{Disjoint boundaries.}\quad
Given two circular boundaries at infinity, we can always find the unique line
perpendicular
to the corresponding hyperplanes in the bulk. If we adjust the cylindrical
coordinates to this axis, the circular boundaries are represented by two circles
${\rho=\infty}$, ${\zeta=\pm\zeta_{\infty}}$. It is possible to find
\cite{KrtousZelnikov:minsurf} a tube-like minimal surface joining these two
boundaries, see Fig.~\ref{fig:ms}a. It is described by the function ${\zeta(P)}$
with ${P=\ch\rho}$, and it is parametrized by ${P_0=\ch\rho_0}$, where
${\rho_0}$ is the closest approach of the surface to the axis:\footnote{The
solutions are expressed in terms of eliptic integrals with the convention of
\cite{GradshteinRyzhik:book}.}
\begin{equation}\label{rotsym-sol}
\begin{split}
    \zeta(P) &{=}\frac{\pm P_0}{\sqrt{1{+}P_0^2}\sqrt{1{+}2P_0^2}}
    \Bigl[
    (1{+}P_0^2)\,\mathsf{F}\Bigl({\textstyle\arccos\frac{P_0}{P},\sqrt{\frac{1+P_0^2}{1+2P_0^2}}}\Bigr)\\
    &\mspace{60mu}
    -P_0^2\, \mathsf{\Pi}\Bigl({\textstyle\arccos\frac{P_0}{P},\frac1{1+P_0^2},\sqrt{\frac{1+P_0^2}{1+2P_0^2}}}\Bigr)
    \Bigr]\;.
\end{split}\raisetag{4ex}
\end{equation}
Setting ${P=\infty}$ we can read out the coordinates ${\pm\zeta_\infty}$ of the circular boundaries:
\begin{equation}\label{rotsym-bound}
\begin{split}
    &\zeta_\infty(P_0) = \frac{P_0}{\sqrt{1+P_0^2}\sqrt{1+2P_0^2}}\\
    &\quad\times
    \Bigl[
    (1+P_0^2)\,\mathsf{K}\Bigl({\textstyle\sqrt{\frac{1+P_0^2}{1+2P_0^2}}}\Bigr)
    -P_0^2\, \mathsf{\Pi}\Bigl({\textstyle\frac1{1+P_0^2},\sqrt{\frac{1+P_0^2}{1+2P_0^2}}}\Bigr)
    \Bigr]\;.
\end{split}\raisetag{8ex}
\end{equation}
The distance between both boundaries, ${s=2\ell\zeta_\infty}$, as a function of
the parameter ${P_0}$ is depicted in Fig.~\ref{fig:area}a. It reveals that the
tube exists only for distances smaller than the maximal distance
${s_\mx\approx1.00229\ell}$, and for these small distances there actually exist
two tube-like minimal surfaces, one shallow one, remaining at large
distances from the axis, and a deep one, approaching the axis. If the distance
of circular boundaries is enlarged, the tube tears off and the minimal surface
discontinuously splits into two trivial hyperplanes spanned on both boundaries.

To estimate which surface is the smallest one, we have to write down the
regularized area:
\begin{equation}\label{rotsym-area}
\begin{split}
  A(P)&=\frac{4\pi\ell^2 P_0^2}{\sqrt{1+2P_0^2}}\,
      \mathsf{\Pi}\Bigl({\textstyle\arccos\frac{P_0}{P},1,\sqrt{\frac{1+P_0^2}{1+2P_0^2}}}\Bigr)\\
    &=2A_\plane + A_\ren +\mathcal{O}\Bigl(\frac1{P^3}\Bigr)\;.
\end{split}
\end{equation}
The divergent term ${A_\plane}$ is given by \eqref{rotsym-planearea}, the finite part ${A_\ren}$ reads
\begin{equation}\label{rotsym-renarea}
  \frac{A_\ren}{4\pi\ell^2} = {\textstyle 1
    {+}{\textstyle\frac{P_0^2}{\sqrt{1{+}2P_0^2}}}
    \mathsf{K}\Bigl({\textstyle\!\sqrt{\frac{1{+}P_0^2}{1{+}2P_0^2}}}\Bigr)
    {-}\sqrt{1{+}2P_0^2}\,\mathsf{E}\Bigl({\textstyle\!\sqrt{\!\frac{1{+}P_0^2}{1{+}2P_0^2}}}\Bigr)
    }\,.
\end{equation}

The renormalized area as a function of ${P_0}$ or of the distance ${s}$ is shown in Fig.~\ref{fig:area}. We see that the shallow tube has always smaller area than the deeper one. However, for ${s_\crit<s<s_\mx}$, the renormalized area of the tube is positive, i.e., the tube has larger area than the trivial solutions of two hyperplanes. The tube is thus the smallest minimal surface only for ${s<s_\crit\approx0.876895\ell}$.

All these exact results confirm the previously conjectured properties based on numerical and approximate analysis \cite{Hirata:2006jx}.


\bigskip\noindent\textbf{Crossing circular boundaries.}\quad
In the case of two crossing circular boundaries at infinity we naturally adjust
the cylindrical coordinates to the axis going through the intersection points.
Thus, the semicircles between these intersection points are represented by lines
${\rho=\infty}$, ${\ph=\text{const}}$. In Poincar{\'e} coordinates they are
half-lines in the plane ${\zP=0}$ starting at ${\rP=0}$. The minimal surface
spanned on two such semicircles is depicted in Fig.~\ref{fig:ms}b. Its explicit
form can be found in \cite{KrtousZelnikov:minsurf}. It exists for any angle
${\phi}$ between both semicircles and can be parametrized  by ${Z_0=\ch\rho_0}$
with ${\rho_0}$ corresponding to the closest approach of the surface to the
axis. The relation between of ${\phi}$ and ${Z_0}$ is one-to-one
\cite{KrtousZelnikov:minsurf}.  The regularized area takes form:
\begin{equation}\label{trsym-area}
\begin{split}
    A(Z)
    &=\frac{2L\ell\, Z_0^2}{Z\sqrt{2Z_0^2-1}}\,
      \mathsf{\Pi}\Bigl({\textstyle\arccos\frac{Z_0}{Z},1,\sqrt{\frac{Z_0^2-1}{2Z_0^2-1}}}\Bigr)\\
    &= A_\plane + \Delta\zeta_*\ell\Bigl[a_\ren+\mathcal{O}\Bigl(\frac1{Z^3}\Bigr)\Bigr]\;.
\end{split}
\end{equation}
The leading term ${A_\plane}$ is given by \eqref{trsym-planearea}. It is
divergent
because of both UV and IR divergences. The next term is proportional to the IR
cut-off ${\Delta\zeta_*}$. The reason is that the minimal surface is invariant
under the translation along the axis. However, we can write down the finite
renormalized area density
${a_\ren=\frac{A_\ren}{\Delta\zeta_*\ell}}$:
\begin{equation}\label{trsym-regarea}
    a_\ren
    {=}2\ell\Bigl[
    {\textstyle\frac{Z_0^2}{\sqrt{2Z_0^2{-}1}}}\,
    \mathsf{K}\Bigl({\textstyle\!\sqrt{\frac{Z_0^2{-}1}{2Z_0^2{-}1}}}\Bigr)
    {-}\sqrt{2Z_0^2{-}1}\,\mathsf{E}\Bigl({\textstyle\!\sqrt{\frac{Z_0^2{-}1}{2Z_0^2{-}1}}}\Bigr)
    \Bigr]\,.
\end{equation}
It is always negative. Naturally, the surface has always smaller area then
two half-hyperplanes starting at the axis reaching the semi-circles at infinity.

\bigskip\noindent\textbf{Touching circular boundaries}\quad
Tangent circular boundaries are trivially represented in Poincar{\'e}
coordinates. If oriented in the ${\yP}$ direction, they are given by ${\zP=0}$,
${\xP=\pm\xP_\infty}$. The minimal surface spanned on such a `strip'
is in Fig.~\ref{fig:ms}c, \cite{Tonni:2010pv,KrtousZelnikov:minsurf}.
It reaches the maximal value of the coordinate ${\zP}$ for ${\zP_0=\xP_\infty/X_0}$,
which we call the `top-line' of the surface. The constant ${X_0}$ is given by
${X_0=\frac{\Gamma(3/4)^2}{\sqrt{2\pi}}\approx 0.59907}$. The area regularized
at ${\zP\ll1}$ is
\begin{equation}\label{horsym-area}
\begin{split}
    A(\zP)
    &=\frac{2\mathcal{A}\ell^2}{\zP_0}\Biggl[
    \sqrt{\frac{\zP_0^2}{\zP^2}-\frac{\zP^2}{\zP_0^2}}
    -\sqrt2\,\mathsf{E}\Bigl({\arccos\frac{\zP}{\zP_0},\frac1{\sqrt2}}\Bigr)\\
    &\mspace{160mu}
    +\frac1{\sqrt2}\,\mathsf{F}\Bigl({\arccos\frac{\zP}{\zP_0},\frac1{\sqrt2}}\Bigr)
    \Biggr]\\
    &=A_\plane +  L_0 \bigl[-2X_0\ell + O(\zP^3)\bigr]\,.
\end{split}\raisetag{3ex}
\end{equation}
The leading divergent term ${A_\plane}$ is given by \eqref{horsym-plane}.
The next term is IR divergent since the surface has the horocyclic symmetry
${\yP\to\yP+\yP_s}$. It is thus proportional to the length
${L_0=\frac{\Delta\yP_*\ell}{\zP_0}}$ measured on the `top-line' of the surface.
The renormalized area density
${a_\ren}={\frac{A_\ren}{L_0}}={-2X_0\ell}$ is, as expected, a constant
independent of the position of the touching circular boundaries.

\section*{Discussion}
\label{sc:summary}

Returning to the conjecture \eqref{Ryu-Takayanagi}, we
can now associate entanglement entropy with any two generally positioned
spherical domains at infinity.
The most interesting case occurs for two disjoint domains. For boundaries
closer than ${s_\mx}$ there are three possible minimal surfaces,
which corresponds to three possibilities (phases) for the holographic
entanglement entropy in CFT.
The physical choice would correspond to the surface of the smallest area.
Inspecting
Fig.~\ref{fig:area}b, one can see that the transition between these phases occurs
at the
distance ${s=s_\crit}$, when the area of the tube-like surface
starts to exceed the area of the trivial solution with two hyperplanes.

If we accept that the entanglement entropy for disjoint subsystems is given
by the absolute minimal surface according to \eqref{Ryu-Takayanagi},\footnote{%
See \cite{Hubeny:2007re} for alternative proposals.}
then the renormalized area \eqref{rotsym-renarea} is directly related to the
mutual information
${I(\Omega_1,\Omega_2)=S_{\Omega_1}+S_{\Omega_2}-S_{\Omega_1\cup\Omega_2}}$
which quantifies correlations between the disjoint subsystems. Indeed, since the
entanglement entropy ${S_\Omega}$ of a single spherical domain ${\Omega}$ is
given by the area ${A_\plane}$ of the trivial hyperplane
boundary ${\partial\Omega}$,
, the renormalized area ${A_\ren}$ of the tube joining the boundaries of two
such
domains gives directly the mutual information ${I(\Omega_1,\Omega_2)}$, provided
that the tube does give the minimal area, i.e., for ${s<s_\crit}$.

Although the entanglement entropy changes continuously with the distance between the
boundaries at ${s=s_\crit}$, the corresponding minimal surface changes discontinuously.
To move from the trivial phase
to the tube-like phase continuously, one would have to start with two very close hyperplanes.
At a point, where they almost touch, a very deep tube-like surface can appear. By enlarging the distance of the boundaries, the tube starts to grow wider. It
follows the upper branch of the curve in Fig.~\ref{fig:area}c (i.e. the
non-physical phase) up to the maximal possible distance ${s_\mx}$ of the
boundaries. Here, one has to start decreasing the distance of the boundaries in
such a way that the tube grows even wider (following the lower branch in Fig.~\ref{fig:area}c). After decreasing the distance under
${s_\crit}$ one obtains, in a continuous way, the physical tube-like phase.

\begin{figure}
\includegraphics[width=1.55in]{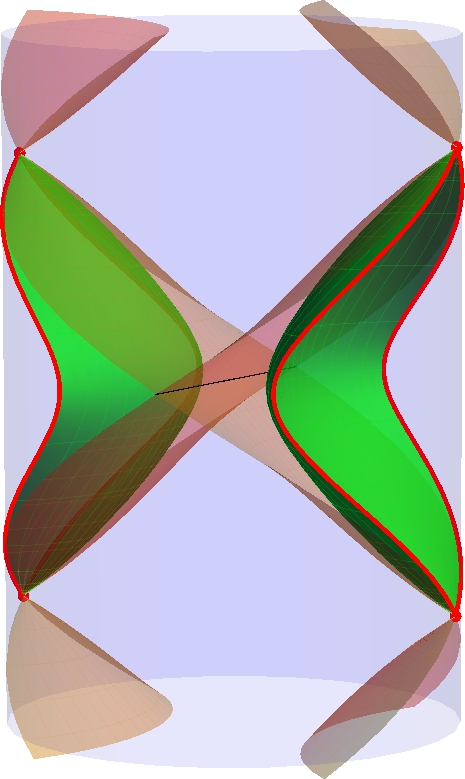}\hfill
\includegraphics[width=1.55in]{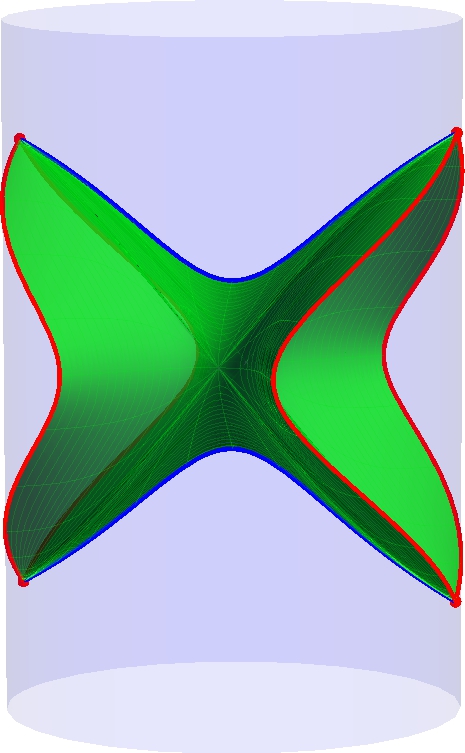}\\[1ex]
{\small (a)\hspace{1.6in}(b)}\\
\caption{\label{fig:AdS}%
\textbf{World-sheets of minimal surfaces.}
The vertical cylinder represents a 3+1 dimensional AdS spacetime with the
angular direction ${\ph}$ suppressed. The world-sheet of one circular boundary
is thus
reduced only to two curves at infinity. One circular boundary is localized in
the left static region, the other in the right one.
(a) Two uniformly accelerated
hyperplanes spanned on these circular boundaries. Killing horizons are
indicated.
(b) The world-sheet of the tube-like minimal surface joining the same circular boundaries.}
\end{figure}

The fact that the tube-like minimal surface does not exist for too distant
boundaries can be explored also in a dynamical way. Although we consider only
static situations---calculation of the minimal surface area given in a
spatial section of a static region of the AdS spacetime---we can take advantage
of the rich structure of AdS symmetries and investigate the situation which
looks rather dynamical in the global picture. Let us consider a static
Killing vector with orbits that have an acceleration larger than ${1/\ell}$.
This Killing vector has a bifurcation character similar to the boost Killing
vector in the Minkowski spacetime. Its Killing horizons divide the AdS space into pairs
of static regions positioned acausally with respect to each other, with
non-static regions between, cf.~Fig.~\ref{fig:AdS}a. The hyperbolic space in
which we found the tube-like solution is a spatial section of both opposite static
regions. We can position one circular boundary at infinity of one static
region and the other one at infinity of the opposite static region. The world-sheets of
the corresponding  hyperplanes describe uniformly accelerated motion along
the Killing vector, see Fig.~\ref{fig:AdS}a. The tube-like minimal surface
can be also evolved into both static regions. However, it does reach the
Killing horizons and there it must be extended into non-static regions.
The resulting surface is depicted in Fig.~\ref{fig:AdS}b.

We see that the surface is non-smooth along two spatial edges, one
describing the formation of the surface in the past, the other its termination in the
future. If the surface is viewed from the perspective of a globally static observer
(the vertical direction in the figure), the future edge can be interpreted as
a tear-off line for  the boundaries positioned too far from each other and the subsequent motion of the separate pieces. For more details, see
\cite{KrtousZelnikov:minsurf}.

Beside the case of two spherical domains we can investigate even more
complicated situations: let us consider spherical domains ${\Omega_i}$,
each of them a subdomain of all the subsequent ones:
${\Omega_i\subset\Omega_j}$ for ${i<j}$. They do not have to be all
simultaneously concentric. The circular boundaries of these domains correspond
to ultraparallel hyperplanes in the bulk.
For such a configuration we know the minimal surfaces for any pair of the
boundaries. Employing \eqref{Ryu-Takayanagi} we find that the renormalized entropy
depends only on the distance of the boundaries, cf.~\eqref{rotsym-bound}, \eqref{rotsym-area}. We can thus test properties of the entropy for domains
obtained by a combination of several subdomains. Namely, one can check
the strong subadditivity inequalities to find that they are satisfied, as
expected from general considerations \cite{Hirata:2006jx}.

Similarly one can study systems of strips between several semicircles joined
at the same poles.

Summarizing, we have found exact analytical solutions for minimal
surfaces in AdS for two disjoint domains at infinity. These classical geometrical solutions
reveal the existence of different phases that reflect a phase transition in
the corresponding quantum CFT, similar to the confinement/deconfinement phase
transition at a finite temperature \cite{Klebanov:2007ws,Lewkowycz:2012mw}. The
holographic entanglement entropy becomes an effective tool for testing phase
transitions in CFT. Note that calculations in purely classical gravity provide an
insight to non-trivial quantum properties of the corresponding field theories.

\section*{Acknowledgments}

P.~K. was supported by Grant GA\v{C}R P203/12/0118
and appreciates the hospitality of the TPI of the University of Alberta.
A.~Z. thanks the Natural Sciences and Engineering
Research Council of Canada and the Killam Trust for
financial support and appreciates the hospitality
of the ITP FMP of Charles University in Prague.
The authors thank Don Page and Martin \v{Z}ofka
for usefull comments on the paper.




%

\end{document}